# Metadamping in inertially amplified metamaterials: Trade-off between spatial attenuation and temporal attenuation

**Mahmoud I. Hussein***
Ann and HJ Smead Department of Aerospace Engineering Sciences,
University of Colorado Boulder, Boulder, Colorado 80303, USA and
Department of Physics, University of Colorado Boulder, Boulder, Colorado 80302, USA

**Ibrahim Patrick**
College of Engineering, Swansea University, Swansea, UK

**Arnab Banerjee**
Department of Civil Engineering, Indian Institute of Technology Delhi, India

**Sondipon Adhikari**
James Watt School of Engineering, The University of Glasgow, Glasgow, UK

Metadamping is the phenomenon of either enhanced or diminished intrinsic dissipation in a material stemming from the material's internal structural dynamics. It has been previously shown that a locally resonant elastic metamaterial may be designed to exhibit higher or lower dissipation compared to a statically equivalent phononic crystal with the same amount of prescribed damping. Here we reveal that even further dissipation, or alternatively further reduction of loss, may be reached in an inertially amplified metamaterial that is also statically equivalent and has the same amount of prescribed damping. This is demonstrated by a passive configuration whereby an attenuation peak is generated by the motion of a mass supported by an inclined lever arm. We further show that by coupling this inertially amplified attenuation peak with that of a local resonance attenuation peak, a trade-off between the intensity of spatial attenuation versus temporal attenuation is realized for a range of the inclination angles. Design for performance along this trade-off is therefore possible by adjustment of the lever angle. These findings open the way for highly expanding the Ashby space for stiffness-damping capacity or stiffness-spatial attenuation capacity through design of the internal structure of materials.

## I. INTRODUCTION

Control of dissipation is one of the key design factors in structural dynamics. Dissipation is a measure of loss of energy as a function of time. In a finite structure, the dissipation characteristics have a spatial and frequency dependency. In an infinite medium, such as a waveguide or a structured material, dissipation exhibits also a wave number (or wave vector) dependency [1-4]. Representation of dissipation as an intrinsic wave number-dependent quantity provides a fundamental measure that is independent of global dimensions or boundary conditions. This measure is obtained by considering a representative unit cell of the medium of interest, applying Bloch's theorem on the unit cell, and calculating the wave number-dependent damping ratio alongside the wave number-dependent damped frequency. Upon integration of each damping ratio branch over the Brillouin zone (BZ), and possibly summing over all branches, a total cumulative measure of dissipation is obtained—which provides a measure of the overall damping capacity of the medium or material under consideration [5].

Phononic materials, such as phononic crystals and elastic metamaterials, provide an opportunity for unit cell design not only to generate band gaps for spatial attenuation but also to control the level of dissipation [6-8]. In phononic crystals, the band structure is shaped by interferences of transmitted and reflected waves from periodic inclusions, interfaces, and/or boundaries within the medium [9{11]. In an elastic metamaterial, which is often also periodic, the band structure is shaped not only by wave interefences but also by couplings—or hybridizations—between substructure resonance modes and elastic wave modes in the hosting medium [12, 13]. Upon comparing the dissipation across the BZ for an elastic metamaterial and a statically equivalent phononic crystal, it was shown that the former may be designed to enable enhanced dissipation, a concept termed metadmping [5, 14]. In some cases, the elastic metamaterial unit cell may be designed to exhibit diminished dissipation, which is negative metadamping [15{18]. Since its inception in 2013 [5], metadamping has been explored in a variety of configurations, including negative-stiffness [19] and non-local resonance [17], and, recently, in materials with electrically activated local resonances [20, 21].

In this work, we investigate metadamping in inertially amplified (IA) metamaterials [22-25]. Inertial amplification in some aspects is an extension of local resonance except that a mechanical mechanism is introduced to provide a magnification of the "effective inertia" of the resonator. This concept may be realized using a lever-arm effect that allows the inertia of a resonating mass to be magnified to a degree proportional to the arm length. We show that an inertially amplified metamaterial may be passively tuned to exhibit a significant further boost in the damping capacity (positive metadamping) or in the reduction of loss (negative metadamping), compared to a statically equivalent locally resonant elastic metamaterial. We examine this behavior in a 1D chain model that encompasses both an IA mass and a local resonance mass separately attached to the baseline mass [26]. This configuration yields a two degrees-of-freedom system with a bounded band gap that features two coupled resonances. In addition to extreme metadamping behavior, this configuration also allows us to reveal a novel regime whereby a trade-off between the intensities of temporal attenuation (dissipation) and spatial attenuation takes effect. These findings have far-reaching implications on the design of future phononic materials with tailored space-time attenuation characteristics.

## II. INERTIALLY AMPLIFIED CHAIN: CONFIGURATION AND MATHEMATICAL MODEL

We consider a one-dimensional infinite chain where the unit cell consists of a base mass $M$ connected by a spring with stiffness $K_{IA}$ and a viscous damping dashpot with damping constant $C$, where the spring and damper act in parallel. Each baseline mass is also connected to its neighbor by an inertial amplifier attachment which comprises an auxiliary mass $m_a$, referred to as the mass of the inertial amplifier, connected to a pair of baseline masses by rigid links. This inertial amplifier mass plays the key role of inducing inertial amplification, as its acceleration is amplified owing to the lever-based connecting mechanism. A single degree-of-freedom linear spring-mass resonator is attached to the baseline mass in the unit cell. The mass and spring constant of the resonator are $m$ and $k$, respectively. The resonator's modal degree of freedom couples with the modal degree of freedom associated with inertially amplified mass and creates a band gap with a double-attenuation peak in the imaginary part of the complex dispersion diagram; this aspect is discussed and analyzed in the next sections. The equation of motion of the resonating mass is

$$m\ddot{w}_x + c(\dot{w}_x - \dot{u}_n) + k(w_x - u_n) = 0, \qquad (1)$$

where $u_n$ and $w_x$ denote the displacement of the baseline mass and the resonator mass, respectively.

*A. Force on the baseline mass from inertial amplifier*

The relationship between the acceleration of the auxiliary masses $\ddot{v}_n$ and the acceleration of the main mass $\ddot{u}_n$ and $\ddot{u}_{n-1}$ are expressed as

$$\ddot{v}_n = \frac{(\ddot{u}_n - \ddot{u}_{n-1})}{2} \cot\alpha, \qquad (2)$$

where α is the angle between the axis of the central axis of the chain and the inertial amplifier mass. The net force on the rigid links is calculated by balancing the forces at the mass of the inertial amplifier as

$$2F_n \sin\alpha = m_a \ddot{v}_n = \big(m_a(\ddot{u}_n - \ddot{u}_{n-1})\big)\frac{\cot\alpha}{2}, \qquad (3)$$

where

$$F_n = \frac{m_a(\ddot{u}_n - \ddot{u}_{n-1})}{4(\sin\alpha)(\tan\alpha)}, \qquad (4)$$

The component of the force acting from inertial amplifier onto the baseline mass along the direction of the wave propagation is

$$\tilde{F}_n = F_n \cos\alpha = \frac{1}{\chi}(m_a(\ddot{u}_n - \ddot{u}_{n-1})), \qquad (5)$$

where $\chi = 4\tan^2\alpha$. From Bloch's theorem,

$$u_{n+1} = u_n e^{i\mu} u_{n-1} = u_n e^{-i\mu}, \qquad (6)$$

where $a$ is the length of the lattice unit, $q$ is the wave number, and a dimensionless term $qa$ is represented by µ.

## B. Equation of motion of the overall chain

The governing equations of motion for the $n^{th}$ baseline mass of the overall inertial amplifier chain are as follows:

where

$$M\ddot{u}_n + c(\dot{u}_n - \dot{w}_x) + k(u_n - w_x) + C(2\dot{u}_n - \dot{u}_{n-1} - \dot{u}_{n+1}) + K_{IA}(2u_n - u_{n-1} - u_{n+1})$$
$$+ \chi(m_a(\ddot{u}_n - \ddot{u}_{n-1})) - \chi(m_a(\ddot{u}_{n+1} - \ddot{u}_n)) = 0 \qquad (7)$$

and

$$[M + 2\chi m_a(1 - \cos\mu)]\ddot{u}_n + [c + 2C(1 - \cos\mu)]\dot{u}_n + [k + 2K_{IA}(1 - \cos\mu)]u_n - c\dot{w}_x - kw_x = 0, \qquad (8)$$

where $\chi = 1/(4\tan^2\alpha)$. In matrix form, we get

$$\begin{bmatrix} M + \frac{1}{2}m_a(1-\cos\mu)\cot^2\alpha & 0 \\ 0 & m \end{bmatrix} \begin{Bmatrix} \ddot{u}_n \\ \ddot{w}_x \end{Bmatrix} + \begin{bmatrix} c + 2C(1-\cos\mu) & -c \\ -c & c \end{bmatrix} \begin{Bmatrix} \dot{u}_n \\ \dot{w}_x \end{Bmatrix} +$$
$$\begin{bmatrix} K + 2K_{IA}(1-\cos\mu) & -k \\ -k & k \end{bmatrix} \begin{Bmatrix} u_n \\ w_x \end{Bmatrix} = \begin{Bmatrix} 0 \\ 0 \end{Bmatrix}, \qquad (9)$$

The reader may refer to Ref. [26] for an effective-dynamical properties analysis of a similar coupled IA-local resonance chain configuration.

To characterize the metadamping performance of the IA chain, we will also consider a statically equivalent phononic crystal (PnC) and a statically equivalent locally resonant acoustic metamaterial (AM). The governing equations for the PnC are

$$m_1\ddot{u}_n + (C+c)\dot{u}_n - c\dot{w}_n - C\dot{w}_{n-1} + (K_{PnC} + k_{PnC})u_n - k_{PnC}w_n - K_{PnC}w_{n-1} = 0, \qquad (10)$$

$$m_1\ddot{w}_n + (C+c)\dot{w}_n - c\dot{u}_n - C\dot{u}_{n+1} + (K_{PnC} + k_{PnC})w_n - k_{PnC}u_n - K_{PnC}u_{n+1} = 0, \qquad (11)$$

and the governing equations for the AM are

$$m_1\ddot{u}_n + C(2\dot{u}_n - \dot{u}_{n-1} - \dot{u}_{n+1}) + c(\dot{u}_n - \dot{w}_n) + K_{AM}(2u_n - u_{n-1} - u_{n+1}) + k_{AM}(u_n - w_n) = 0, \qquad (12)$$

$$m_2\ddot{w}_n + c(\dot{w}_n - \dot{u}_n) + K_{AM}((w_n - u_n) = 0. \qquad (13)$$

We will first consider the IA chain with the local resonator mass set as zero, i.e., $m = 0$. In this case, for the PnC and AM, we will assume $m_1 = M$ and $m_2 = m_a$. We will then consider a nonzero local resonator mass and assume $m_1 = M + m_a$ and $m_2 = m$.

## III. EXTREME LEVELS OF POSITIVE AND NEGATIVE METADAMPING

We first investigate an inertially amplified chain without a resonating mass, for which the unit cell is depicted in the third, fourth, and fifth panels in Fig. 1a for three different IA angles, $\alpha = 13, 24,$ and 43, respectively. For comparison, a standard diatomic phononic crystal model and a mass-in-mass acoustic metamaterial model, each with the same long-wave speed of sound as the IA chain, are considered. This provides us with five distinct chains that are all statically equivalent. The parameters used for all models are given in Table I. The damped frequency dispersion diagrams and corresponding damping ratio diagrams for each of the five systems are given in Figs. 1b and 1c, respectively. In all frequency band structure figures, we plot the quantity $\Omega = \omega/\omega_b$, where $\omega_b = K_{PnC}/M$. The complex dispersion for the undamped version of all models are also provided to show the frequency-dependent profile of spatial attenuation for each chain. In Fig. 1d, the wave number-dependent damping emergence metric, $Z$, which is a measure of metadamping is plotted. This metric represents the difference between the damping ratio of the AM or each of the IA chains and that of the PnC. Any increase in $Z(\mu)$ above zero is indicative of positive metadamping, and, in contrast, a decrease in $Z(\mu)$ below zero is indicative of negative metadamping. The quantity with "sum" in the superscript represents the integrated value of $Z$. Upon complete integration over the Brillouin zone, a total value of $Z$, with "tot" in the supersciprt is calculated to give an overall quantification of the positive or negative damping capacity with respect to the reference PnC. Since the IA chain here has only one branch, the $Z$ quantity in Fig. 1d is obtained by comparing the damping ratio of that branch with the average of the two damping ratio branches of the PnC. Alternatively, in Fig. 1e, the comparison is made with the damping ratio of the first branch of the PnC. It is observed from the results that for the parameters selected, as listed in Table 1, the IA exhibits positive metadamping with increasing levels as $\alpha$ increases. In contrast, the AM chain is showing a mild level of negative metadamping.

TABLE I. Parameters used for the unit-cell models examined in Figs. 1 and 2

| Parameter | Figure 1 | Figure 2 |
|---|---|---|
| $M$ | 1 | 1 |
| $m$ | - | 0.1 |
| $m_a$ | 0.1 | 0.1 |
| $K_{PnC}$ | 120 | 120 |
| $k_{PnC}$ | 1320 | 1200 |
| $K_{AM}$ | 110 | 109 |
| $k_{AM}$ | 179 | 179 |
| $K_{IA}$ | 100 | 100 |
| $k_{IA}$ | - | 40 |
| $C$ | 2 | 2 |
| $c$ | 0 | 2 |

We now attach a local resonator to the base of the IA chain, as shown in the third, fourth, and fifth panel schematics of Fig. 2a. This allows us to consider models with the same number of dispersion branches, and, importantly, gives us an IA configuration that may exhibit a bounded band gap, as would an experimentally realizable material modeled as a continuum. An undamped version of this model has been recently investigated for spatial attenuation characteristics [26]. Here we observe a monotonic increase in positive metadamping, as expressed by the quantity $Z_{sum}^{tot}$, as we progress from a AM chain to a IA chain with increasing values of the lever-arm angle $\alpha$. This demonstrates an extreme level of positive metadamping significantly exceeding what is realized by an AM. Figure 3 extends the analysis of both models, the inertially amplified chain without (Fig. 3a) or with (Fig. 3b) a local resonator to a broad range of quasi-static speeds extending well beyond the values considered in Figs. 1 and 2; this provides an Ashby-like map for the damping capacity versus the long-wave speed, which is representative of the effective quasi-static stiffness of the chain. We observe in Fig. 3a that the AM exhibits positive metadamping throughout the parameter range shown. We also observe that the inertially amplified chain (without a resonator) exhibits even more positive metamping for $\alpha = 43$; however, the level of metadamping decreases as $\alpha$ decreases. Eventually, for $\alpha = 13$, the metadamping is shown to be strongly negative, i.e., the IA chain exhibits less loss overall compared to the statically equivalent PnC. In Fig. 3b, where a local resonator is attached to the IA chain, we observe positive metadamping with increasing intensity as level-arm $\alpha$ increases.

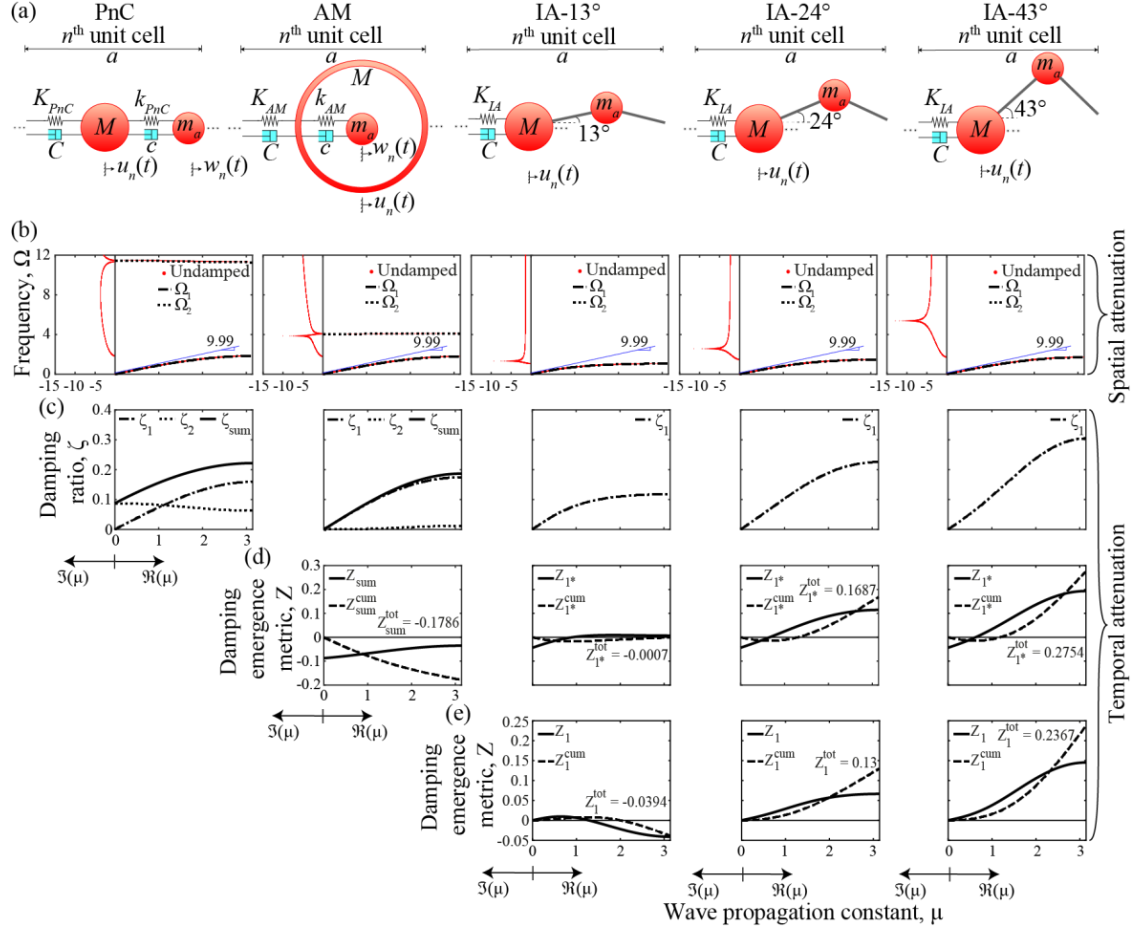

FIG. 1. Analysis of inertially amplified chain without a resonating mass. (a) Unit-cell schematics of statically equivalent chains, including a PnC, AM, and three variations of the inertially amplified chain exhibiting $\alpha = 13$, 24, and 43, respectively. All IA masses are restricted to motion in only the vertical direction. (b) Frequency band structure for the undamped and damped unit-cells; results for the damped unit cells only shown for real wave numbers. (c) Damping-ratio diagrams for the damped unit cells. (d) Damping-emergence metric $Z$; for the IA, the average of the two damping ratio branches of the PnC has been used as a reference. (e) Damping-emergence metric $Z$ for the IA where damping ratio corresponding to the acoustic branch of the PnC is used as a reference. In sub-figures (b) and (c), the subscripts '1' and '2', and 'sum' indicate the acoustic branch, optical branch, and sum of the two branches, respectively.

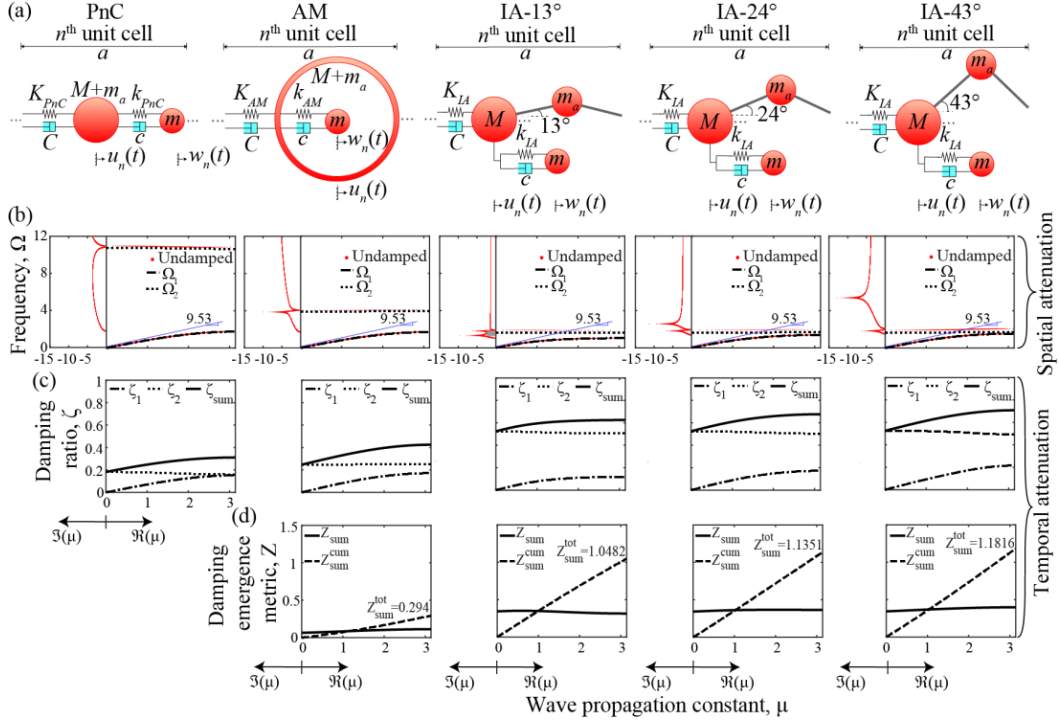

FIG. 2. Analysis of inertially amplified chain with a resonating mass. (a) Unit-cell schematics of statically equivalent chains, including a PnC, AM, and three variations of the inertially amplified chain exhibiting $\alpha = 13$, 24, and 43, respectively. All IA masses are restricted to motion in only the vertical direction. (b) Frequency band structure for the undamped and damped unit-cells; results for the damped unit cells only shown for real wave numbers. (c) Damping-ratio diagrams for the damped unit cells. (d) Damping-emergence metric $Z$. In subfigures (a) and (b), the subscripts '1' and '2', and 'sum' indicate the acoustic branch, optical branch, and sum of the two branches, respectively.

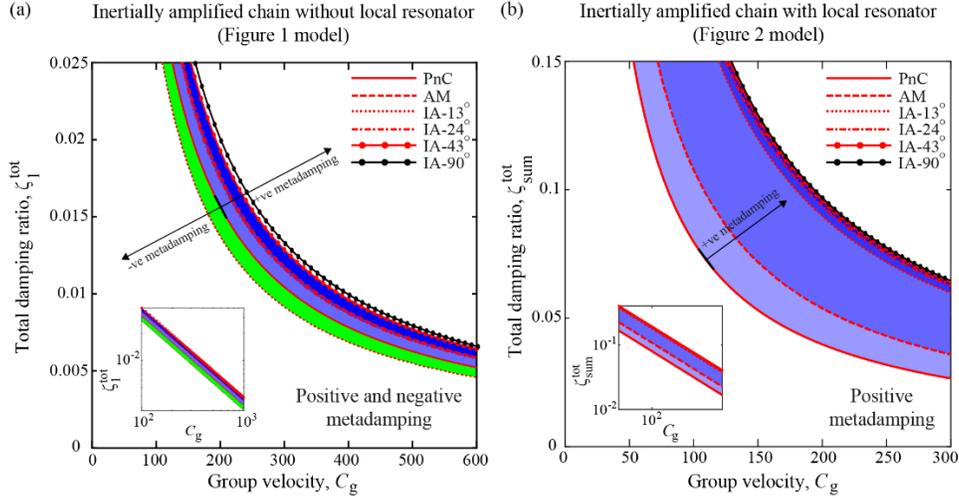

FIG. 3. Illustration of metadamping: total damping ratio based on the first branch $\zeta_1^{tot}$ or the sum of the two branches $\zeta_{sum}^{tot}$ versus long-wave speed of sound $C_g$ in the periodic chains. (a) Positive or negative metadamping (depending on value of $\alpha$) exhibited in the case of inertially amplified chain without a resonating mass; blue- and green-shaded areas represent regions of positive and negative metadamping, respectively. (b) Positive metadamping exhibited in the case of inertially amplified chain with a resonating mass; blue-shaded area represents region of positive metadamping. The intensity of positive metadamping increases as the colour of the blue shaded areas gets darker. The insets depict the corresponding plots in a log-log scale.

## IV. REGIME DISPLAYING TRADE-OFF BETWEEN TEMPORAL ATTENUATION AND SPATIAL ATTENUATION

In Fig. 4, we expand our examination of the IA chain with a local resonator and further examine the effects of the level-arm angle $\alpha$ on the dissipation, i.e., temporal attenuation. However, we also examine its effect on the imaginary wavenumber part of the spectrum, which represent the spatial attenuation. As shown in [26], for a certain range of $\alpha$, the attenuation peak associated with IA mechanism couples with the attenuation peak associated with the local resonator. This yields a coupled double-attenuation peak, within a bounded band gap, in the imaginary wavenumber part of the dispersion diagram. We can quantify the strength of this attenuation by evaluating the minimum value of imaginary wave propagation constant bounded by the two peaks (see Ref. [26]).

Using these measures, Figs. 4a and 4b respectively show the variation of the temporal and spatial attenuation with the IA level arm $\alpha$. Figure 4c combines these two relations in one plot, showing the temporal attenuation versus the spatial attenuation. It is observed that the two measures increase monotonically up to a specific angle, $\alpha = 17.532$. However, as the angle increases further, a trade-off is observed between the intensity of temporal attenuation and the intensity of spatial attenuation. To the best of the authors' knowledge, this represents an unprecedented space-time attenuation trade-off phenomenon for an elastodynamic medium. This trade-off is observed until $\alpha = 18.348$, beyond which the coupled double-peak feature in the spatial attenuation spectrum no longer exists, and hence the $\mu_{\min}$ measure is no longer valid. Figure 4d shows the frequency and damping ratio diagrams for a selection of points in the parameter space, as marked in Figs. 4c.

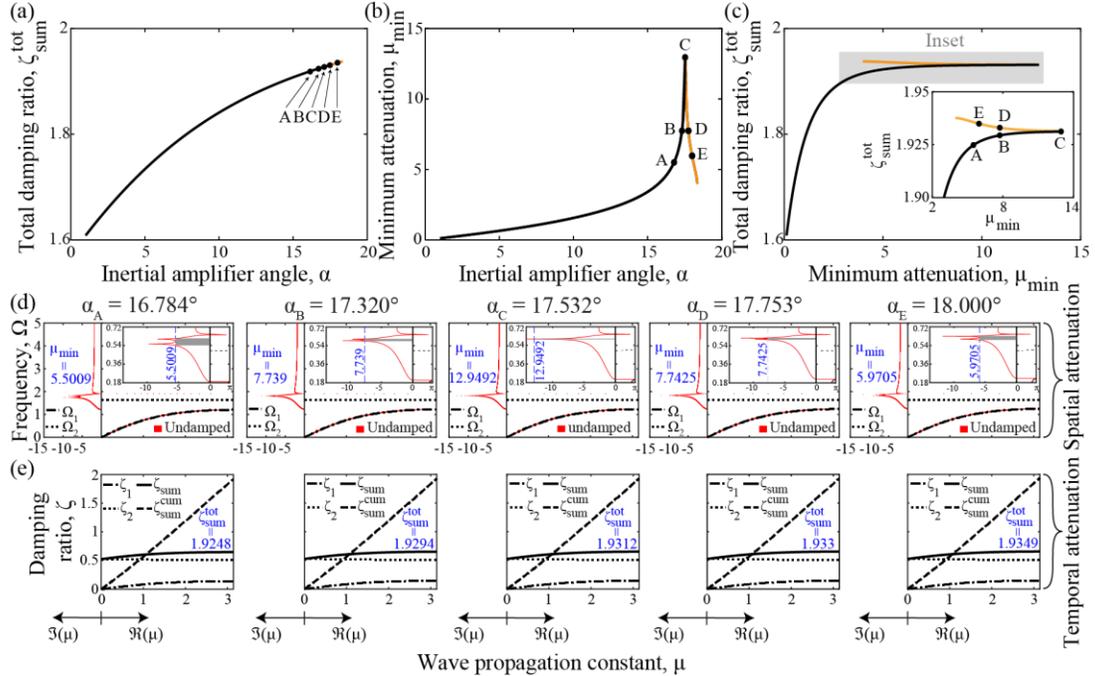

FIG. 4. Performance characteristics for the inertially amplified chain with a resonating mass. (a) Temporal attenuation: total damping ratio $\zeta_{\text{sum}}^{\text{tot}}$ versus inertial amplifier angle $\alpha$. (b) Spatial attenuation: minimum attenuation, $\mu\min$ versus $\alpha$. (c) The quantities $\mu_{\min}$ versus $\zeta_{\text{sum}}^{\text{tot}}$. The inset, in sub-figure (c), shows an enlarged version of the trade-off region. The points A-E, corresponding to five different values for $\alpha$, have been specifically chosen and used to illustrate the evolution of the response with the angle $\alpha$ where a region is identified that features a trade-off between the temporal and spatial attenuation properties. (d) Frequency band structure corresponding to the five chosen values of $\alpha$; insets show the corresponding plots with the y-axis reproduced in a log scale and results for the damped IA shown only for real wave numbers. (e) Damping-ratio diagrams corresponding to the five chosen values of $\alpha$.

## V. CONCLUSIONS

We have demonstrated that inertial amplification provides a route to yielding a structured material with simultaneously high stiffness and high damping capacity (positive metadamping) or low stiffness and low loss (negative metadamping). The levels of positive or negative metadamping attained are elevated, well exceeding the performance of conventional acoustic metamaterials. These attributes extend the boundaries of viscoelastic dynamical properties of state-of-the-art structured material systems. Furthermore, when combining the inertial amplification with local resonance, we observe the unique phenomenon of a trade-off between temporal attenuation and spatial attenuation intensities in the material properties. This is realized passively and by only changing the lever angle in the IA chain. This trait opens the way for the design of future phononic materials with tailored space-time attenuation characteristics and could have significant implications on topological phononics and other contemporary areas in phonon engineering.